\documentclass[pra,twocolumn,showpacs,preprintnumbers,amsmath,amssymb]{revtex4}
\usepackage{mathrsfs}
\usepackage{bbm}
\usepackage{amssymb}
\usepackage{amsfonts}
\usepackage{tipa}
\usepackage[colorlinks,linkcolor=blue,hyperindex,CJKbookmarks]{hyperref}
\usepackage{epsfig,graphicx}
\usepackage{amstext}
\usepackage{amsmath}
\usepackage{graphicx}
\newcommand{\Det}{\textmd{Det} }

\begin{document}

\title{Suppressing correlated noise in signals transmitted over the Gaussian memory channels using $2N$-port splitter and phase flips}

\author{Ke-Xia Jiang$^{1,2,}$}
\email{kexiajiang@126.com}
\author{Shi-Quan Zhang$^{2,}$}
\author{San-Min Ke$^{3,}$}
\author{Heng Fan$^{1,}$
}
\email{hfan@iphy.ac.cn}
\address{$^{1}$Beijing National Laboratory for Condensed Matter Physics,
               Institute of Physics, Chinese Academy of Sciences, Beijing
               100190, P. R. China \\
         $^{2}$Department of Physics, Engineering University of CAPF, Xi'an 710086, P. R. China\\
         $^{3}$College of Science, Chang'an  University, Xi'an 710064, P. R. China}
\begin{abstract}
A scheme for suppressing the correlated noise in signals transmitted over the bosonic Gaussian
memory channels is proposed. This is a compromise solution rather than removing the noise completely.
The scheme is based on linear optical elements, two $N$-port splitters and $N$ number of phase flips.
The proposed scheme has the advantages that the correlated noise of the memory channels are greatly suppressed,
and the input signal states can be protected excellently when transmitting over the noise channels.
We examine the suppressing efficiency of the scheme for the correlated noise, both from
quantum information of the states directly transmitted through
the noise channel and also from the entanglement teleportation.
The operation of phase flips in our scheme is important
for the suppression of the correlated noise, which
can diminish the effect of noise in quantum communication.
Increasing the number of beam splitters also can improve the suppressing efficiency of the scheme
in quantum communication.

\end{abstract}
\maketitle
\section{Introduction}\label{sec.Introduction}
Quantum information generally encoded in a time ordered
sequence of quantum states like photons can be transmitted from Alice to Bob through a quantum channel.
Quantum state transfer can be performed by local quantum operations with assistance
of classical communication by teleportation \cite{Bennett1993}.
The success of teleportation relies on the pre-shared entanglement as the resource
whose construction also needs quantum states transmitted through quantum channels.
For noiseless quantum channel, quantum information transfer by both
schemes of flying qubits and teleportation can be implemented ideally
without the effect of decoherence. This fact of decoherence free is also
important for quantum key distribution protocols which can provide
unconditional secure quantum communication \cite{Gisin2002}.
However, the performance of realistic
communication channel will be limited
by noises. In noisy quantum communication, the input signals are
contaminated by the inevitable external environment interactions.
This may induce a significant information loss and hence reduce also the communication security.
Protecting the information against noise contamination is one of the most important tasks
for quantum information processing. We remark that quantum information
can be both in discrete or continuous variable (CV) systems \cite{cvrmp,cvprotocol}.

A channel is called memoryless if
the noise acts identically and independently on
each element of the sequence.
For memoryless channels,
one standard strategy of protecting quantum information
is to use quantum error-correcting codes~\cite{qec}.
This strategy gives detailed guidance for
encoding and decoding in communication procedures.
However, in current communication systems, considerable channel with noise correlated in time
and space is existing with the increasing speed of optical communication and the miniaturization
of solid state~\cite{Bowen2005,Kretschmann2005,Corney2006}. For example, when the typical
environmental relaxation times are comparable with the time delays between two signals,
the channels present correlations or memory.
These scenarios are called correlated noise channels or memory channels~\cite{Kretschmann2005,Caruso2012}.
The efficacy of standard quantum error-correcting
codes is reduced substantially, when signals are transmitted over such memory channels~\cite{memorycodes}.
The technical problem is finding the optimal encoding schemes,
which is rather complex and only a few models have been solved~\cite{Kretschmann2005,Datta2007,Arrigo2007,Giovannetti2005}.

Very recently, some attentions have been devoted to consider new encoding-decoding procedures
for protecting signal states against the correlated noise in bosonic quantum channels
both theoretically~\cite{Lupo2012} and experimentally~\cite{Lassen2013}.
By introducing unitaries prepended and appended to the memory channel,
the removal of the correlated noise can be implemented partially in Ref.~\cite{Lupo2012}.
However, accurately definition of the unitaries have
challenges for the composition of elementary gates.
In Ref.~\cite{Lassen2013}, an encoding and
decoding scheme was proposed for protecting the input single quantum states near ideally.
The encoding technique is based on two main steps:
combining the input signal with auxiliary vacuum
states on a beam splitter and subsequently introducing
phase flips for one of the two resulting states. The decoding technique
is inverse similarly.
However, a predetermined condition required:
the magnitude of the correlated noise should be known
for structuring the transmissivity of beam splitters.
Realistically, the magnitude of the correlated noise may not be accurately
grasped when considering the randomly varying of the external environments~\cite{magnitude}.
This limits the execution of encoding-decoding procedures.

In this work, we propose a error-protecting scheme for suppressing
the correlated noise in signals transmitted
over bosonic Gaussian memory channels
based on linear optical elements.
We construct an encoding-decoding procedure by aiding
two $N$-port splitters (the $2N$-port splitter)~\cite{2Nsplitter} and $N$ phase flips,
but no predetermined condition of the correlated noise should be required.
We examine the suppressing efficiency of the correlated noise both from
quantum information of the states directly transmit over
the noise channel and also from the entanglement teleportation.
This is a compromise solution rather removing the noise completely.
The proposed scheme achieves good performance, where
the correlated noise are greatly suppressed and the signal input states are exhibited excellent protections.
We would like to remark that here the encoding is different from the
codewords in quantum error correction~\cite{qec}.

This paper is organized as follows: in Sec.~\ref{sec.scheme}, we introduce our error-protecting scheme
for suppressing the correlated noise in a bosonic Gaussian memory channel generally. In Sec.~\ref{efficiency},
a specific memory model is investigated.
The high suppressing efficiency for the correlated noise is analyzed.
In Sec.~\ref{suppressing}, we examine the suppressing effects
both from directly transmit quantum information over the noisy channel and also from the entanglement teleportation.
Numerical results are illustrated in figures.
The final section, Sec.~\ref{summary}, is devoted to conclusions and discussions.

\section{The scheme for suppressing correlated noise}\label{sec.scheme}
\begin{figure}[htbp]
\centering
\resizebox{0.47\textwidth}{!}{%
\includegraphics{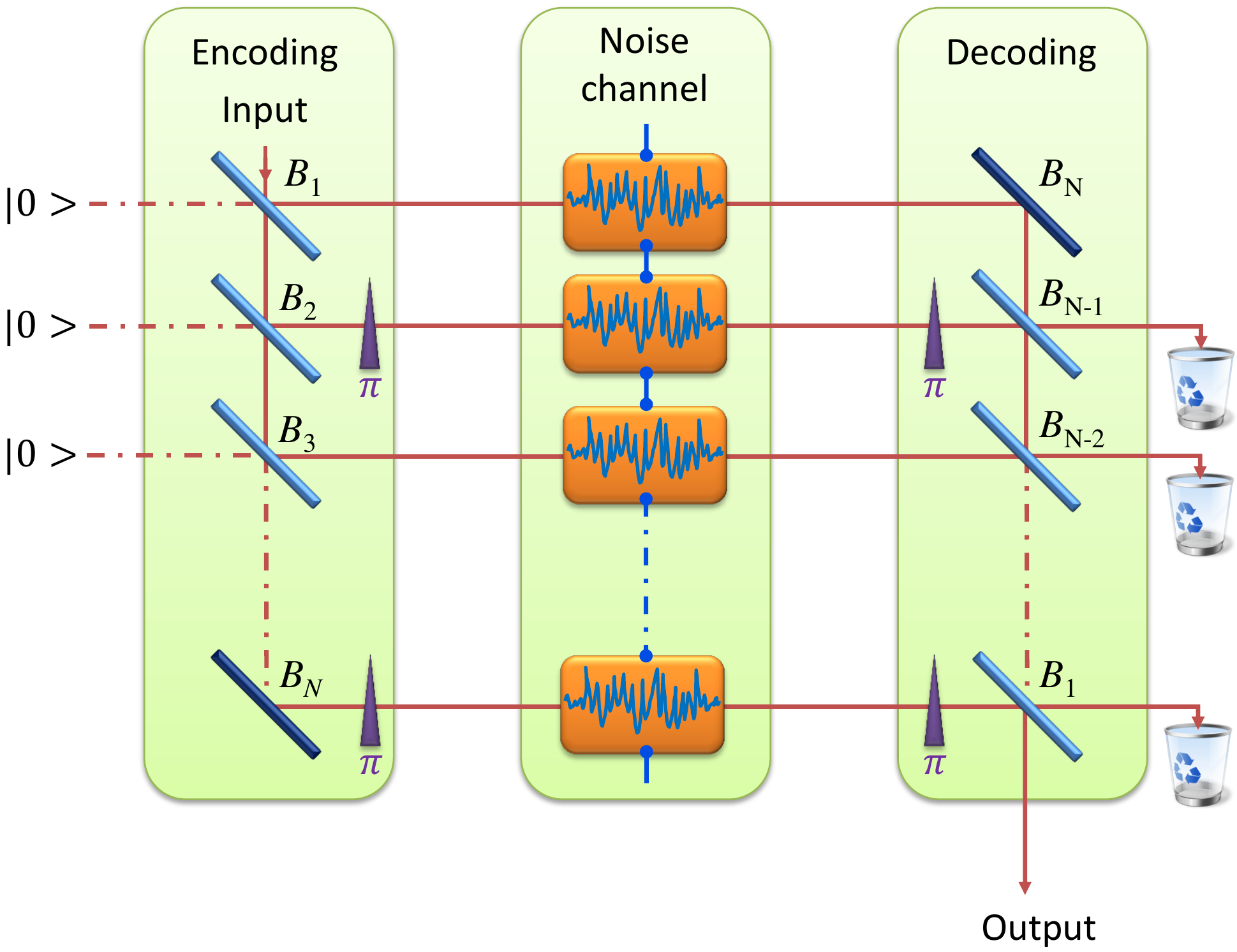}}
\caption{(Color online) The scheme for suppressing correlated noise when signals transmitting over the noise channels.
 Firstly, the input signals are encoded by the first $N$-port splitters and $N/2$ phase flips which implemented
 immediately on the even or odd number of channels. Then, the encoded signals are transmitted over
 the noise channels and contaminated by the noise unavoidably. Finally, the other $N/2$ phase flips and the seconde $N$-port splitters
 are implemented in the decoding procedure. Noise components are being filtered out subtotally and
 the signals are being purified. The short vertical lines between the noise channels denote the correlated noise.
 }\label{scheme}
\end{figure}

Our scheme for suppressing correlated noise in signals transmitted over noise channels
is illustrated in Fig.~\ref{scheme}. The scheme is suitable for depicting both consecutive uses
of a single channel with temporal correlations (e.g. memory channels) and
spatially separated channels with spatial correlations. The encoding and decoding procedures are
realized by the $2N$-port splitter and $N$ phase flips.
The phase flips are implemented after the first $N$-port splitter and
before the seconde (inverse) one, but only on the even (or odd)
number of channels. Auxiliary vacuum states are being added in the encoding procedure.
Finally, in the decoding procedure, the noise components are being
filtered out subtotally and the signals are being purified.
\subsection{The encoding procedure}\label{encodinge}
\begin{figure}[b]
\centering
\begin{minipage}[c]{0.23\textwidth}
\centering
\includegraphics[width=1.5in]{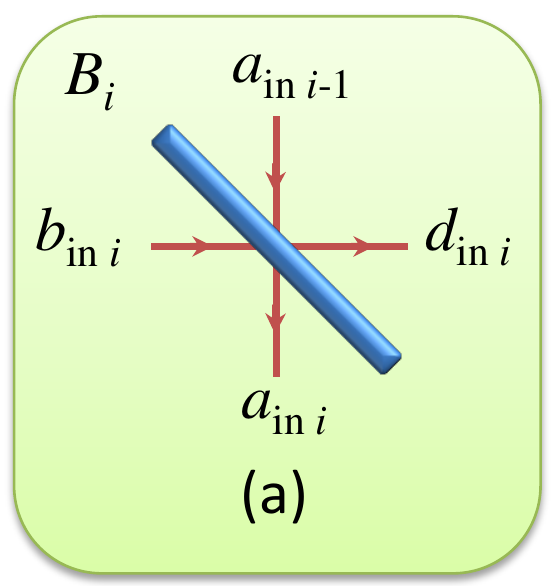}
\end{minipage}%
\begin{minipage}[c]{0.23\textwidth}
\centering
\includegraphics[width=1.5in]{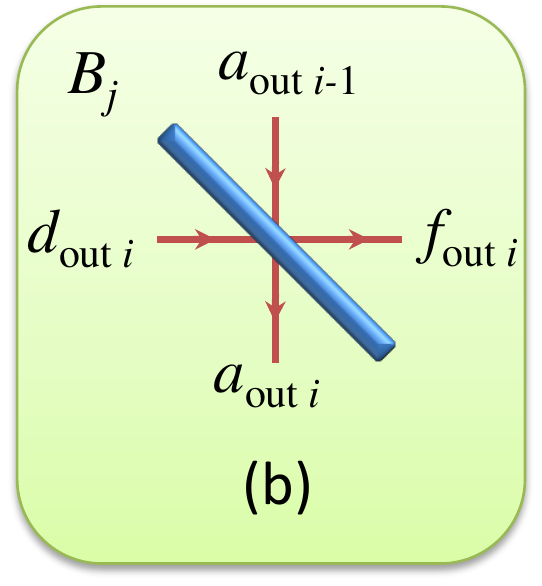}
\end{minipage}
\caption{(Color online)
The splitters in the encoding (a) and decoding (b) procedures. The pairs of input and output
operators can be related by a unitary transformation with the the reflectivity $r_{i}$
and transmissivity $t_{i}$ for both splitters in the encoding and decoding procedures.
}\label{splitters}
\end{figure}
The encoding procedure is realized by aiding an $N$-port splitter $\{B_{i}\}_{i=1,2,\ldots,N}$
and $N/2$ phase flips which implemented immediately on the even (or odd) number of channels.
Without loss of generality, we may assume that $N$ is even in the following analysis.

The pairs of input and output operators on splitters are related by a unitary transformation (see Fig.~\ref{splitters}(a))
\begin{equation}\label{splitterinput}
\begin{aligned}
  d_{\textmd{in}i}&=t_{i} b_{\textmd{in}i}-r_{i} a_{\textmd{in}i-1}, \\
  a_{\textmd{in}i}&=t_{i} a_{\textmd{in}i-1}+r_{i} b_{\textmd{in}i},
\end{aligned}
   \end{equation}
where the reflectivity and transmissivity of the splitter are $r_{i}=\sqrt{\frac{1}{N-i+1}}$ and $t_{i}
=\sqrt{\frac{N-i}{N-i+1}}$, respectively. Operators $b_{\textmd{in}i}$ are responsible for the auxiliary vacuum states.
While the operator $a_{\textmd{in}0}$ corresponds to the input signal states, and
denotes $a_{\textmd{in}}=a_{\textmd{in}0}$ in the following paper.

Using Eq.~\eqref{splitterinput}, the input operators of the memory channels can be rewritten as
\begin{equation}\label{channelinput}
d_{\textmd{in}i}=-\frac{1}{\sqrt{N}}a_{\textmd{in}}-\sum_{j=0}^{i-1} r_{j} r_{j+1} b_{\textmd{in} j}+ t_{i} b_{\textmd{in} i}.
\end{equation}
The relative phase shifts of $\pi$ (namely the phase flips) which being implemented on the even number of channels,
can be described by performing the substitutions
\begin{equation}\label{pfin}
        d_{\textmd{in} i} \mapsto (-1)^{i+1}d_{\textmd{in} i}.
\end{equation}

\subsection{The memory channel}\label{channel}
As an example of the performance of this scheme, we are interesting in
a lossy Bosonic memory channel introduced by
Lupo {\em et al.}~\cite{Lupo2010}. There are two specific highlights of such a mode:
one is the memory kernel, which account for the
flux of information from one channel use to the following,
and other is introducing a parameter \emph{transmissivity} $\epsilon$
to relate the ratio between the time delay $\Delta t$ of the two
successive channel uses and the typical relaxation time $\tau_{rel}$
of the environment, namely $\epsilon\simeq \exp(-\Delta t/\tau_{rel})$.
Conveniently, we call the transmissivity $\epsilon$ the \emph{memory factor}.

In this model, the action of the channel upon the $k$-th
use is defined by a concatenation of $n$ identical unitary
transformations which couples the input modes $\{d_{\textmd{in} j}\}_{j=1,2,\ldots,n}$
with a collection of local environments $\{e_j\}_{j=1,2,\ldots,n}$
and the memory kernel $m_1$. The input signals from different channel use interfere at the
channel outputs, leading to the memory effects of the quantum channel.

Generally, in the Heisenberg picture, the $k$-th output mode $d_{\textmd{out}k}$ can be written as
\begin{equation}\label{memorychannel}
d_{\textmd{out}k}=\sum_{j=1}^{k} f_{k j} d_{\textmd{in} j}+\sum_{j=1}^{k} g_{k j} e_{j}+h_{k} m_{1}.
\end{equation}
The coefficients $f_{k j}$ describe the channels transmittances
and crosstalks between the different use of channels,
$g_{k j}$ describe local environmental contaminations and $h_{k}$ represents
the initial memory components.
The characteristics of the memory channel are reflected by these coefficients.
They are all functions of parameters of the channel transmissivity and the
memory factor. When one takes $f_{k j}=\sqrt{\eta}\delta_{k,j}$,
$g_{k j}=\sqrt{1-\eta}\delta_{k,j}$ and $h_{k}\equiv 0$ with the channel transmissivity $\eta$,
a memory channel can be easily reduced to the memoryless one.

\subsection{The decoding procedure}\label{decodinge}
The decoding procedure is constructed by the other $N/2$ phase flips, which implemented
on the even or odd number of channels, and also the other inverse $N$-port splitters.
The transformation of the input and output operators on the $i$-th splitter satisfies
(see Fig.~\ref{splitters}(b))
\begin{equation}\label{splitteroutput}
\begin{aligned}
  f_{\textmd{out}i}&=t_{j} d_{\textmd{out}i}+r_{j} a_{\textmd{out}i-1}, \\
  a_{\textmd{out}i}&=t_{j} a_{\textmd{out}i-1}-r_{j} d_{\textmd{out}i}.
\end{aligned}
   \end{equation}
We will denote $a_{\textmd{out}}=a_{\textmd{out}N}$
in the following paper without misunderstanding.

From the optical circuit of the scheme in Fig.~\ref{scheme},
it is easy to find that the $i$-th output signal states of the channel
corresponds to  the $j$-th inverse splitter in the decoding procedure,
with the relation $j=N-i+1$.
Using the above equations, one can obtained
the final output of the ladder operators
\begin{equation}\label{channeloutput}
a_{\textmd{out}}=-\frac{1}{\sqrt{N}} \sum_{k=1}^{N} d_{\textmd{out} k}.
\end{equation}
Mathematically, the $N/2$ phase flips can be performed by substitutions:
$d_{\textmd{out} k} \mapsto (-1)^{k+1} d_{\textmd{out} k}$.
After some calculations, we have
\begin{equation}\label{finalout}
    a_{\textmd{out}}=\zeta_{\textmd{in}} a_{in}-\sum_{i=1}^{N-1} \zeta_{\textmd{b} i} b_{\textmd{in} i}
                      +\sum_{i=1}^{N}(-1)^{i+1} \zeta_{\textmd{e} i} e_{i}+\zeta_{\textmd{m}} m_{1},
\end{equation}
with the coefficients
\begin{equation}\label{coefficients}
\begin{aligned}
  \zeta_{\textmd{in}} &=\frac{1}{N} \sum_{k=1}^{N}\sum_{j=1}^{k} (-1)^{j+k} f_{kj},\\
  \zeta_{\textmd{b} i}&=\frac{1}{\sqrt{N}} \Big{[} t_{i}\sum_{j=i}^{N}(-1)^{j+i}f_{ji}
                        -r_{i}r_{i+1} \\
                        &\hspace{30mm}\times \sum_{j=i+1}^{N}\sum_{k=i+1}^{j} (-1)^{j+k} f_{jk}\Big{]},\\
  \zeta_{\textmd{e} i}&=\frac{1}{\sqrt{N}}\sum_{k=j}^{N} (-1)^{j+k+1} g_{kj},\\
  \zeta_{\textmd{m}}  &=\frac{1}{\sqrt{N}}\sum_{k=1}^{N} (-1)^{k} h_{k}.
\end{aligned}
\end{equation}
The coefficients also satisfy the relation of normalization
\begin{equation}\label{coefficientunit}
\zeta_{\textmd{in}}^2+\sum_{i=1}^{N-1}\zeta_{\textmd{b} i}^2+ \sum_{i=1}^{N}\zeta_{\textmd{e} i}^2+\zeta_{\textmd{m}}^2=1.
\end{equation}

These coefficients reflect the complex components of the output signal state.
It is mixed by the auxiliary vacuum states, the
local environmental states and the memory components of the quantum channel.
The magnitude which they contribute can be expressed by the square of coefficients.
However, in our scheme the input signal states have
gotten protections and the correlated noise have been suppressed.
A preliminary understanding can be grasped from the
additions and reductions appear alternately in the Sigma of
the coefficients \eqref{coefficients}.
The coefficients of noises are greatly consumed by the Sigma of additions and reductions.
In the next section we consider a specific model of lossy bosonic memory channel, where
all the coefficients can be calculated numerically.

Phase flip is an important technique to suppress the correlated noise of the memory channels.
If the phase flips are not being implemented, the optical circuit
only simulates the $N$ channel use where the memory effect exist,
but without any suppression of the noise components.
The results of output operators can be easily read out
from Eqs.~\eqref{coefficients}: let all the $j$ and $k$ behind the symbol of Sigma equal to $1$.
In the following sections, in order to demonstrate the high
efficiency of the scheme for suppressing correlated noise,
we will compare the two different results numerically.
\begin{figure*}[htbp]
\centering
\begin{minipage}[c]{0.44\textwidth}
\centering
\includegraphics[width=3in]{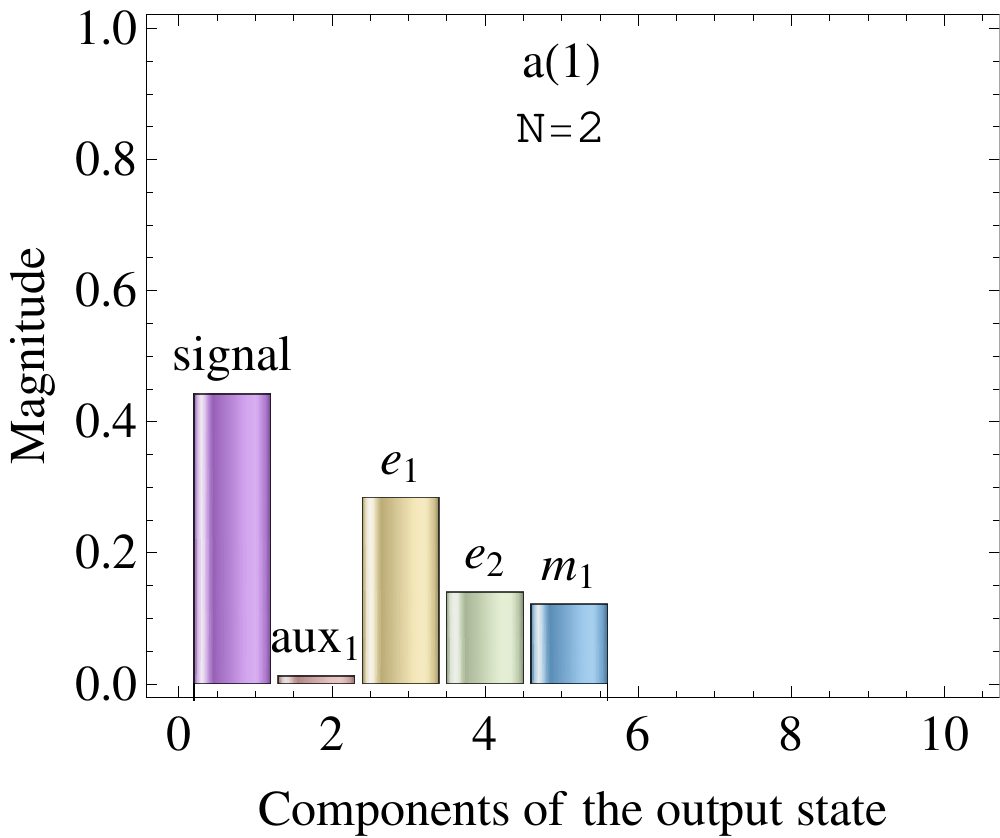}
\end{minipage}
\begin{minipage}[c]{0.44\textwidth}\centering
\includegraphics[width=3in]{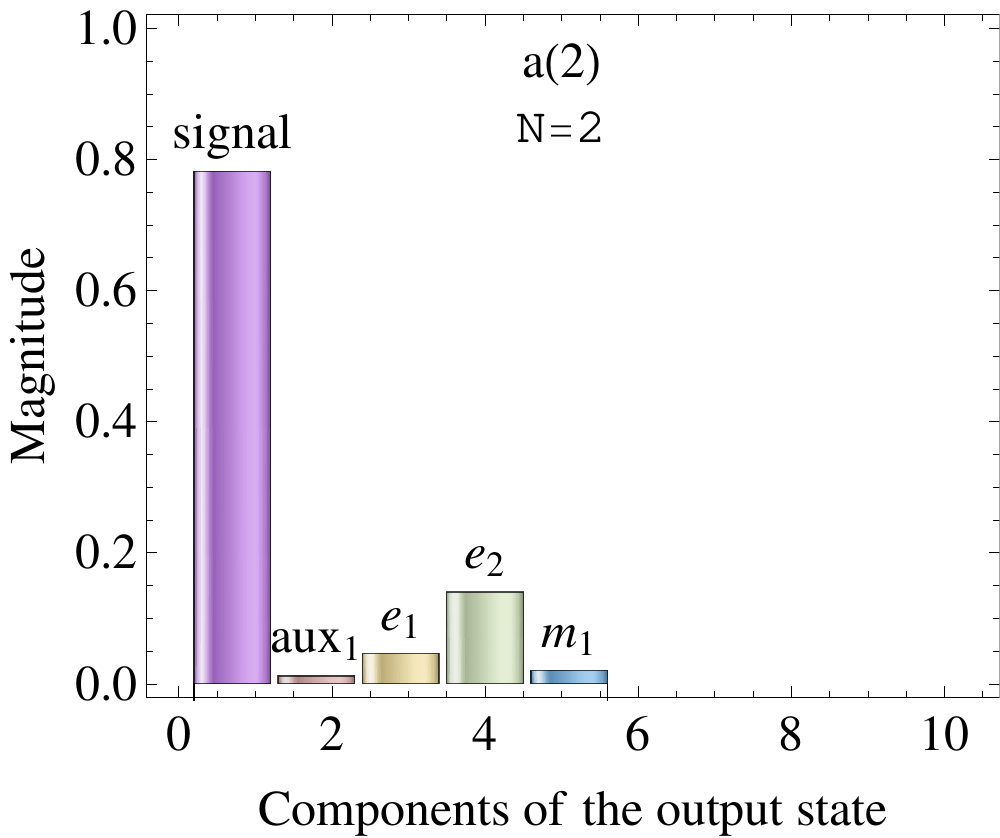}
\end{minipage}\\
\vspace*{.5cm}
\centering
\begin{minipage}[c]{0.44\textwidth}
\centering
\includegraphics[width=3in]{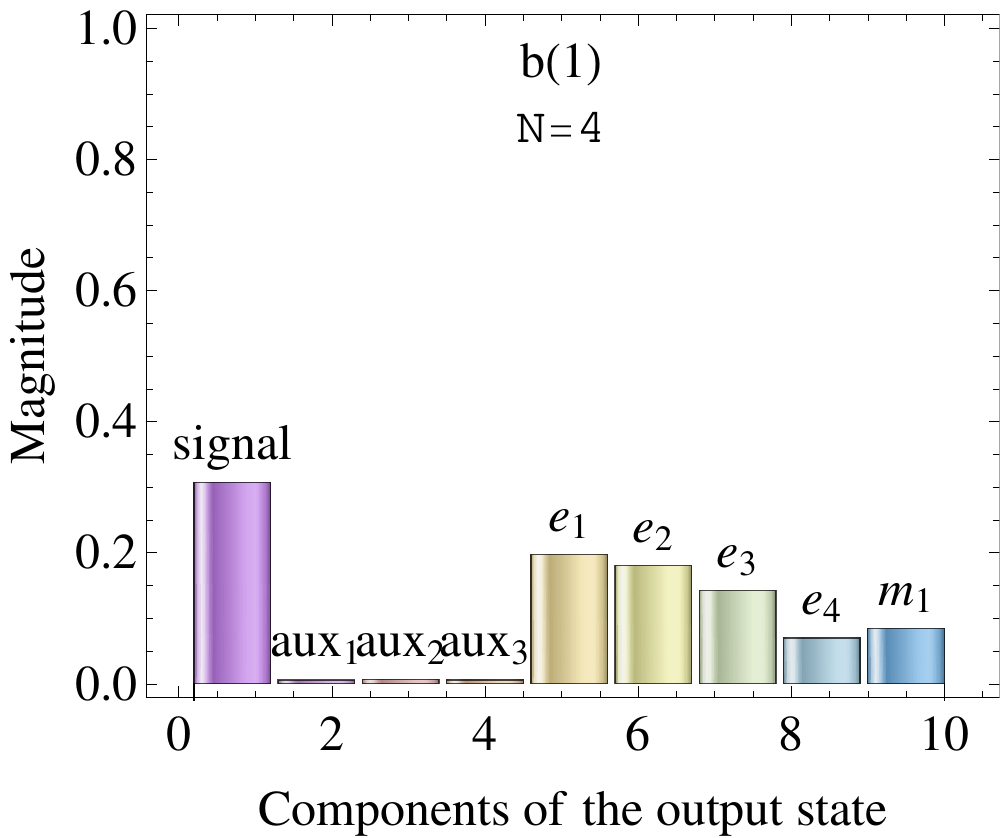}
\end{minipage}
\begin{minipage}[c]{0.44\textwidth}\centering
\includegraphics[width=3in]{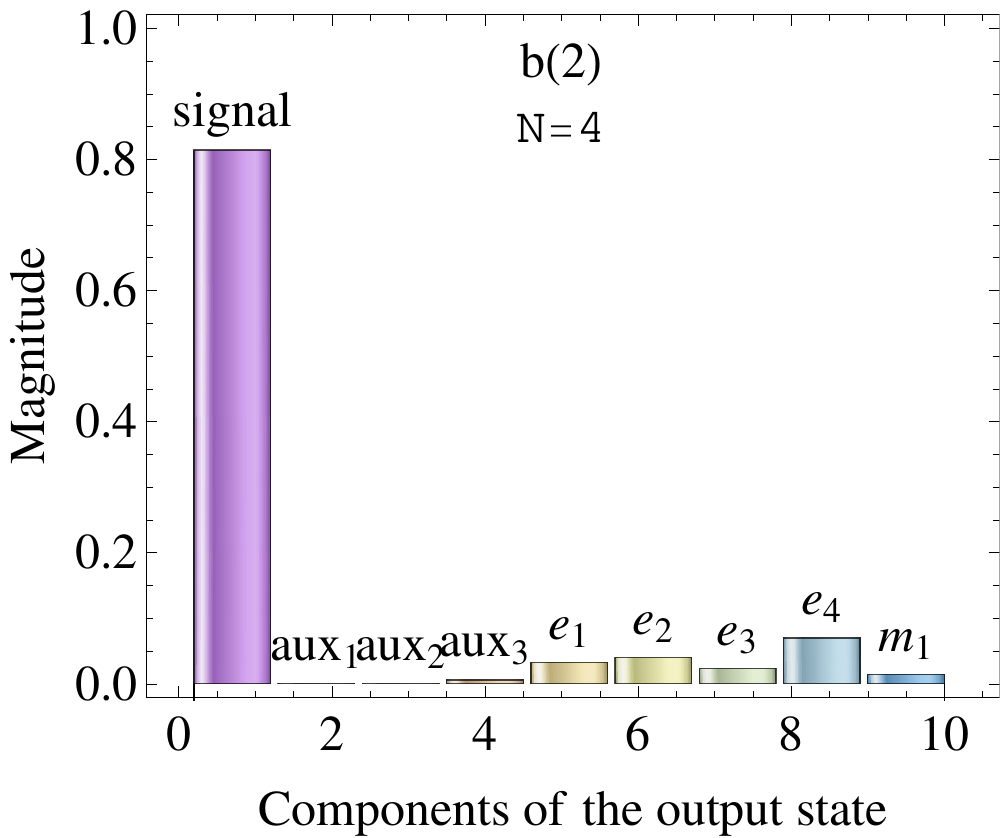}
\end{minipage}
\caption{(Color online)
Bar Charts of the magnitude for components which contribute in the output states.
We have taken the transmissivity $\eta=0.6$ and memory factor $\epsilon=0.3$,
respectively, for the lossy bosonic memory channel Eq.~\eqref{memorycoefficientf}.
Phase flips have not been implemented in figures a(1) and b(1), while in a(2) and b(2) they have been implemented.
The signs aux$_{i}$ express the auxiliary components which correspond to the operators $b_{\textmd{in} i}$.
In the suppressing scheme, when the phase flips have been implemented,
the magnitude of the signal in the output state has greatly been improved, while all the other noise components have been suppressed.
}\label{ratioN24}
\end{figure*}
\section{High efficiency of the signal transmission over lossy bosonic memory channels}\label{efficiency}

In this section we consider a more concrete model of lossy bosonic memory channel which is introduced and
characterized by Lupo \textit{et al}. in \cite{Lupo2010}.
We analyze the suppressing effects of the correlated noise, numerically.

The lossy bosonic memory channel Eq.~\eqref{memorychannel} can be written in a detailed form
\begin{align}\label{memorycoefficientf}
d_{\textmd{out}k} = & \sqrt{\eta} \, d_{\textmd{in}k} - \sqrt{\epsilon}(1-\eta) \sum_{j=1}^{k-1}
                      \left(\sqrt{\epsilon\eta}\right)^{k-j-1} d_{\textmd{in}j} \nonumber\\
                    & -\sqrt{(1-\epsilon)(1-\eta)} \, \sum_{j=1}^k \left(\sqrt{\epsilon\eta}\right)^{k-j} e_{j} \nonumber \\
                    & + \sqrt{\epsilon(1-\eta)}\left(\sqrt{\epsilon\eta}\right)^{k-1} m_1
\end{align}
for $k=1, 2, \dots, N$.
The parameter $\eta$ can be understood as the channel transmissivity of the input signal for a single channel use.
And the parameter $\epsilon$ is the memory factor.

As we have analyzed in the previous section,
the complex components of the output states can be reflected
by the magnitude which they contribute, namely, the square of coefficients Eq.~\eqref{coefficients}.
In Fig.~\ref{ratioN24}, we plot bar charts of the square of coefficients for
ladders of the output states of the memory channel Eq.~\eqref{memorycoefficientf}
with $N=2$ and $N=4$, respectively.
Numerical results show that the phase flips play key roles for the suppression
of correlated noise in the lossy bosonic memory channels.
We compare the two different results numerically for both the phase flips are
being implemented and not implemented.
The magnitude of correlated noises in signals, i.e. $\{e_j\}_{j=1,2,\ldots,n}$
and also the memory kernel $m_1$, are greatly suppressed after the phase flips being implicated.
Correspondingly, the efficiency of the signals are greatly improved.
Interestingly, as depicted in the bar charts of the figures,
it is always greater than the case for a single channel use.

Although from the numerical distributions of the coefficients,
we can see the high efficiency of the scheme for suppressing of the correlated noise,
but a detailed analysis of the signal quantum state protection in communication is also necessary.
Quantum information transmits through the channel can be carried out in two ways:
one is the information directly through the channel, and the other is the teleportation communication
by using entanglements. In the next section,
we will analyze the suppressing effects for the correlated noise of the memory channel from the above two aspects.

\section{The suppressing effects of the input quantum state transmitted over Gaussian memory Channels}\label{suppressing}

The suppressing scheme is universal. It is valid for any input quantum state
and no matter with the statistics of the correlated noise.
In this section, we firstly investigate our scheme
for coherent states in a Gaussian noise environment. And then we investigate the survival
of entanglements in the suppressing scheme. Only let one half of the entangled state sending through the
Gaussian memory channel.
Specifically, as an example, we will use the two-mode squeezed vacuum state (TMSVs) as the entanglement resource.
Since the input signal states are Gaussian states~\cite{Ferraro2005},
the final output states also have Gaussian characteristics when
transmitted over such Gaussian memory channels~\cite{LupoMancini2010}.

We rewrite the Eq.~\eqref{finalout} as
\begin{equation}\label{channela}
    a_{\textmd{out}}=\zeta_{\textmd{in}} a_{\textmd{in}}+X^\mathsf{T} b +Y^\mathsf{T} e+\zeta_{m} m_{1},
\end{equation}
where we denote the operators $b:=(b_{\textmd{in} 1}, \cdots, b_{\textmd{in} N})^\mathsf{T}$ and $e:=(e_1,\dots e_N)^\mathsf{T}$.
The vectors $X$, $Y$ can be read out from Eq.~\eqref{finalout} and Eq.~\eqref{coefficients}, easily.
Furtherly, by using the quadrature operators $q=(a+a^{\dag})/\sqrt{2}$, $p=(a-a^{\dag})/i\sqrt{2}$,
Eq.~\eqref{channela} can be concisely expressed as
\begin{equation}\label{channelR}
    R_{\textmd{out}}=\zeta_{\textmd{in}} R_{\textmd{in}}+\tilde{X}^\mathsf{T} R_\textmd{b} +\tilde{Y}^\mathsf{T} R_e+\zeta_{m} R_{\textmd{m}_1},
\end{equation}
where the operators
$R_{\textmd{out}}:=(q_{\textmd{out}}, p_{\textmd{out}})^\mathsf{T}$,
$R_{\textmd{in}}: =(q_{\textmd{in}}, p_{\textmd{in}})^\mathsf{T}$,
$R_\textmd{b}:=(q_{\textmd{b}_1}, p_{\textmd{b}_1}, \cdots, q_{\textmd{b}_N}, p_{\textmd{b}_N})^\mathsf{T} $,
$R_\textmd{e}:=(q_{\textmd{e}_1}, p_{\textmd{e}_1}, \cdots, q_{\textmd{e}_N}, p_{\textmd{e}_N})^\mathsf{T}$ and
$R_{\textmd{m}_1}:=(q_{\textmd{m}_1}, p_{\textmd{m}_1})^\mathsf{T}$.
The tilde vectors have the forms
\begin{align}\label{tildeXY}
\tilde{X}^\mathsf{T}&=\left(
                                 \begin{array}{cc}
                                   X^\mathsf{T} & 0 \\
                                   0 & X^\mathsf{T} \\
                                 \end{array}
                               \right) \Lambda,\\
\tilde{Y}^\mathsf{T}&=\left(
                                 \begin{array}{cc}
                                   Y^\mathsf{T} & 0 \\
                                   0 & Y^\mathsf{T} \\
                                 \end{array}
                               \right) \Lambda,
\end{align}
with the matrix $\Lambda=(\lambda_{ij})_{2N \times 2N}$ and
\begin{align}\label{matrixLambda}
\lambda_{ij}    =
                  \left\{
                      \begin{aligned}
                        &\delta_{i,2i-1} \hspace{3mm}&& (i\leq N)  \\
                        &\delta_{i,2(i-N)} \hspace{3mm} &&(i\geq N+1)  \\
                      \end{aligned}
                  \right..
\end{align}

The input Gaussian state can be regarded as a combination of the input signal state,
the auxiliary vacuum states, the initial memory mode and the environmental modes. It can be characterized by the first
moments $\langle R_{\textmd{in}} \rangle$, $\langle R_\textmd{b} \rangle$, $\langle R_\textmd{e} \rangle$,
$ \langle R_{\textmd{m}_1} \rangle$,
and the covariance matrix
\begin{eqnarray}\label{inoutCM}
V = \left(\begin{array}{cccc}
            V_\textmd{in} &   0          & 0             &  C^\mathsf{T} \\
               0          & V_\textmd{b} & 0             &  0 \\
               0          & 0            & V_\textmd{e}  &  D^\mathsf{T}\\
               C          & 0            & D             &  V_{\textmd{m}_1}
\end{array}\right),
\end{eqnarray}
where the off-diagonal terms account for possible correlations of the initial memory mode with the input and the environment modes.
Conveniently, we denote $d_{\textmd{in}}=\langle R_{\textmd{in}} \rangle$
in the following paper.
When the Gaussian input state transmitted over the Gaussian channel,
the output state also has Gaussian characterizes, with the first moment~\cite{LupoMancini2010}
\begin{align}\label{first moment}
    d_{\textmd{out}}:&=\langle R_{\textmd{out}} \rangle\nonumber\\
    &=\zeta_{\textmd{in}}^2 \langle R_{\textmd{in}} \rangle
    +\tilde{X}^\mathsf{T} \langle R_\textmd{b} \rangle +\tilde{Y}^\mathsf{T} \langle R_\textmd{e} \rangle +\zeta_{\textmd{m}}^2 \langle R_{\textmd{m}_1} \rangle,
\end{align}
and the covariance matrix
\begin{align}\label{outcovariance}
V_{\textmd{out}}= \zeta_{\textmd{in}}^2 V_{\textmd{in}}&+\zeta_{\textmd{m}}\left[ \zeta_{\textmd{in}}
                   (\tilde{X}^\mathsf{T} C+ C^\mathsf{T} \tilde{X})+\tilde{Y}^\mathsf{T} D+ D^\mathsf{T} \tilde{X} \right]\nonumber\\
                 & +\tilde{X}^\mathsf{T} V_\textmd{b}\tilde{X}+\tilde{Y}^\mathsf{T} V_\textmd{e}\tilde{Y}
                 +\zeta_{\textmd{m}}^2  V_{\textmd{m}_1}.
\end{align}

Generally, a Gaussian channel can be characterized by a triad $(d_{C}, X_{C}, Y_{C})$ \cite{Holevo2001,LupoMancini2010},
which transforms the input Gaussian state $(d_{\textmd{in}}, V_{\textmd{in}})$
to the output Gaussian state $(d_{\textmd{out}}, V_{\textmd{out}})$ as
\begin{align}
    d_{\textmd{out}} &=X_{C} d_{\textmd{in}}+d_{C},\label{triad01}\\
    V_{\textmd{out}} &=X_{C} V_{\textmd{in}}X_{C}^\mathsf{T}+ Y_{C}.\label{triad02}
\end{align}
So it is not difficult to understand that, in this situation, the suppressing scheme can be considered as
a single noise Gaussian channel use, which associates with the matrix
\begin{align}\label{scheme triad}
X_{C}&=\zeta_{\textmd{in}} \mathbb{I},\\
Y_{C}&=\zeta_{\textmd{m}}\left[ \zeta_{\textmd{in}}
                   (\tilde{X}^\mathsf{T} C+ C^\mathsf{T} \tilde{X})+\tilde{Y}^\mathsf{T} D+ D^\mathsf{T} \tilde{X} \right]\nonumber\\
                 &~~~~ +\tilde{X}^\mathsf{T} V_\textmd{b}\tilde{X}+\tilde{Y}^\mathsf{T} V_\textmd{e}\tilde{Y}
                 +\zeta_{\textmd{m}}^2  V_{\textmd{m}_1}.
\end{align}
\begin{figure*}[htbp]
\centering
\begin{minipage}[c]{0.44\textwidth}
\centering
\includegraphics[width=3.3in]{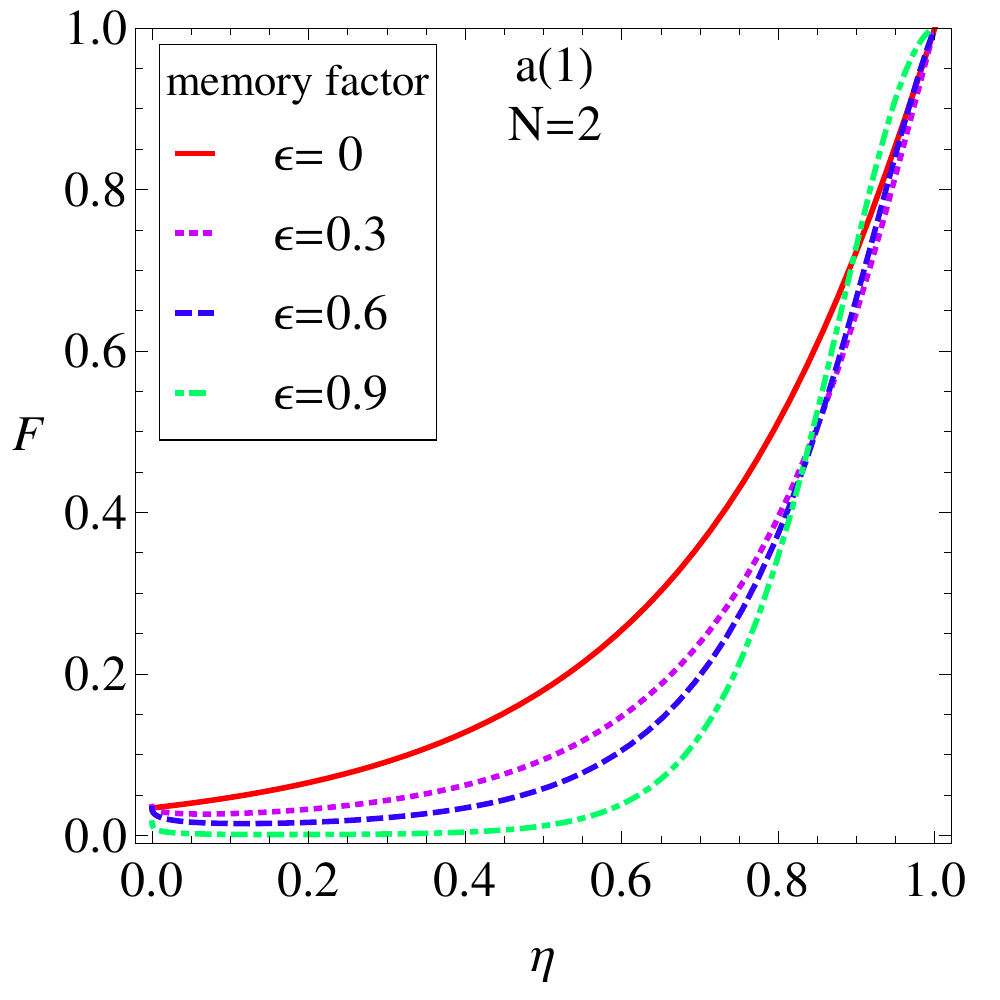}
\end{minipage}
\begin{minipage}[c]{0.44\textwidth}\centering
\includegraphics[width=3.3in]{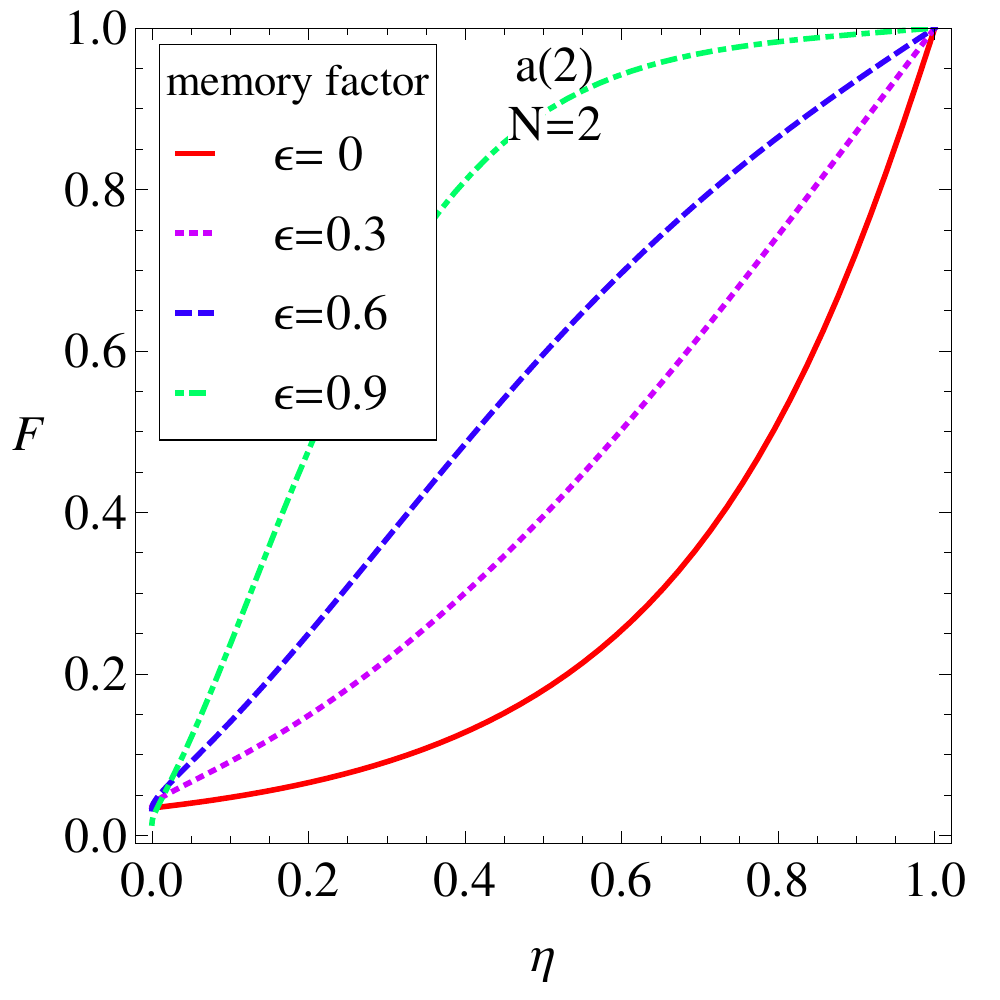}
\end{minipage}\\
\centering
\begin{minipage}[c]{0.44\textwidth}
\centering
\includegraphics[width=3.3in]{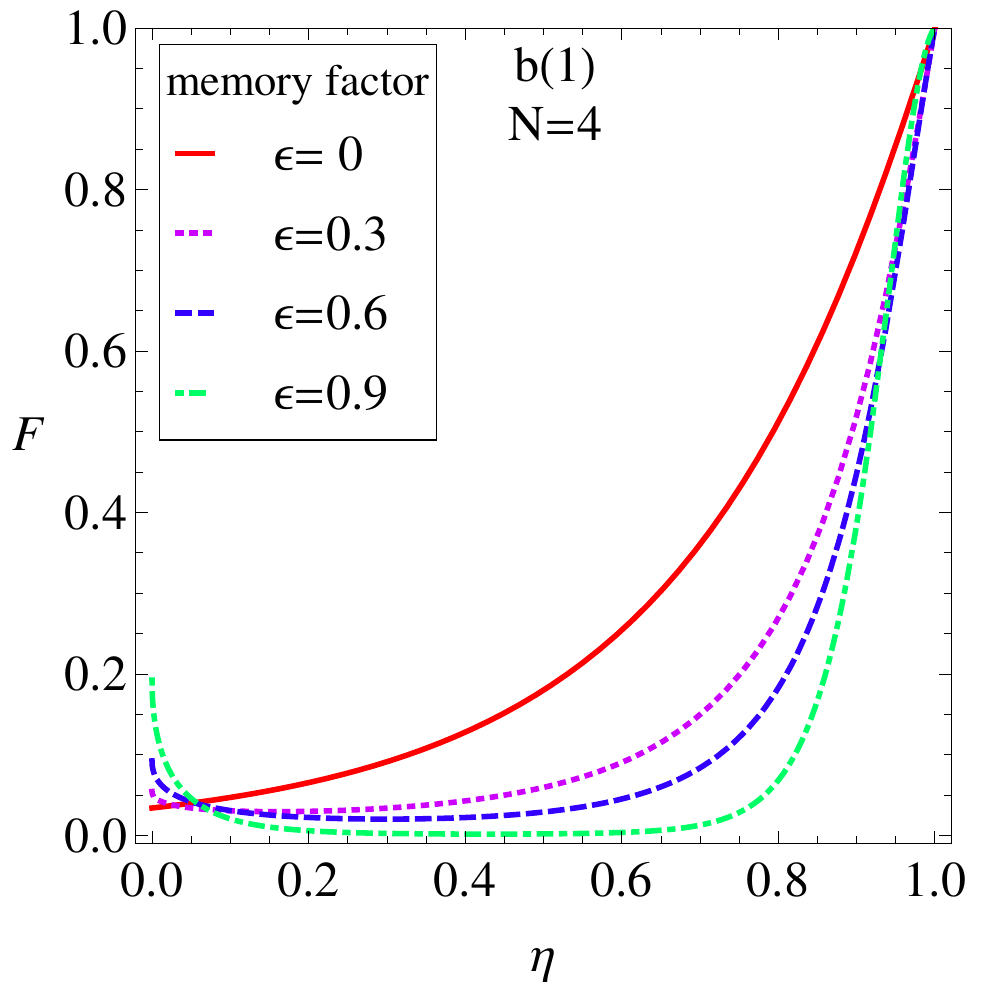}
\end{minipage}
\begin{minipage}[c]{0.44\textwidth}\centering
\includegraphics[width=3.3in]{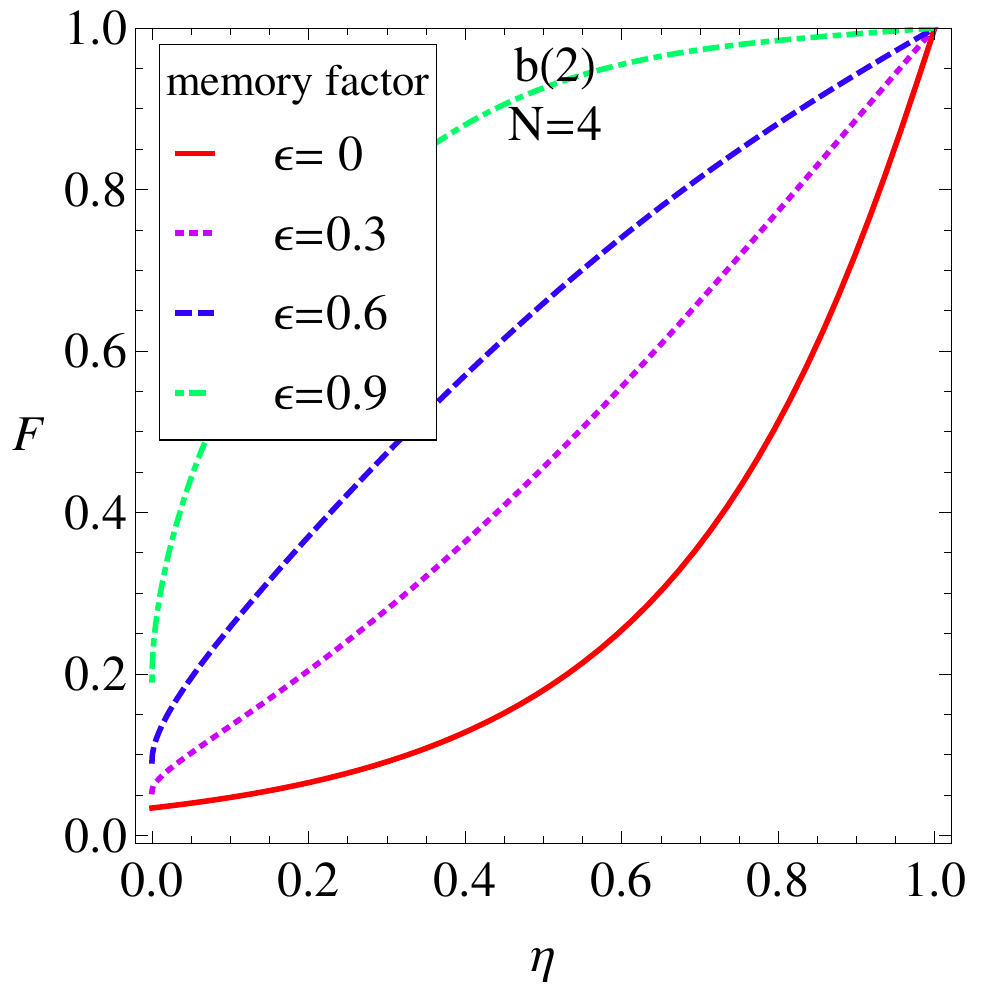}
\end{minipage}
\caption{(Color online)
The fidelity of the input coherent state transmits over a lossy bosonic
Gaussian memory channel in suppressing scheme of the correlated noise.
We have taken the values of the parameters $T=3$ and $|\alpha|^2 =8$
in numerical analysis.  In order to compare the results,
phase flips have been implemented in figures a(2) and b(2),
while in figures a(1) and b(1) they have not been taken.}\label{fidelity ps}
\end{figure*}
\subsection{The fidelity for transmission of coherent states}\label{coherent states}

Considering the teleportation of an ensemble of pure states,
the fidelity, $F=\text{Tr}[\rho_{\text{in}}\rho_{\text{out}}]$, is an appropriate measurement,
which describes how close between the initial input state
and the final (mixed) output quantum state.

For a pure input Gaussian signal state described by $(d_{\textmd{in}},V_{\textmd{in}})$,
the fidelity for transmitting over the general Gaussian channel $(d_{C}, X_{C}, Y_{C})$
can be expressed as \cite{Scutaru1998,Meisheng2007,Isara2008,Marian2012}
\begin{align}\label{Gaussian fidelity}
    F=&\frac{1}{\sqrt{\Det(V_{\textmd{in}}+V_{\textmd{out}})}}\textmd{exp}\bigg\{ -\frac{1}{2}\Big[(X_C-1)d_{\textmd{in}}+d_C\Big]^\mathsf{T}\nonumber\\
    &\times\frac{1}{V_{\textmd{in}}+V_{\textmd{out}}}\Big[(X_C-1)d_{\textmd{in}}+d_C\Big]\bigg\},
\end{align}
where $V_{\textmd{out}}$ can be read out from \eqref{triad02}.
In our paper, we are only interested in the maximum of the fidelity, then we set the channel $d_C = 0$.

In the following analysis, we assume the
initial state of the environment to be the thermal state with the average excitations $T$ for per environmental mode.
While the initial memory and auxiliary modes are vacuum states.
We let the first moment of all the above states values zero,
so as to facilitate analysis and appraisal the suppressing effects of the scheme.
The covariance matrixes have forms~\cite{Ferraro2005}:
$V_\textmd{b}=\mathbb{I}_{2N \times 2N}/2$, $V_\textmd{e}=(T+1/2)\mathbb{I}_{2N \times 2N}$ and
$V_{\textmd{m}_1}=\mathbb{I}_{2 \times 2}/2$.
It is worth noting that all these modes are independent of each other,
so we have $C=0$ and $D=0$ for the off-diagonal terms in \eqref{inoutCM}.

\begin{figure*}[htbp]
\centering
\begin{minipage}[c]{0.45\textwidth}
\centering
\includegraphics[width=3.3in]{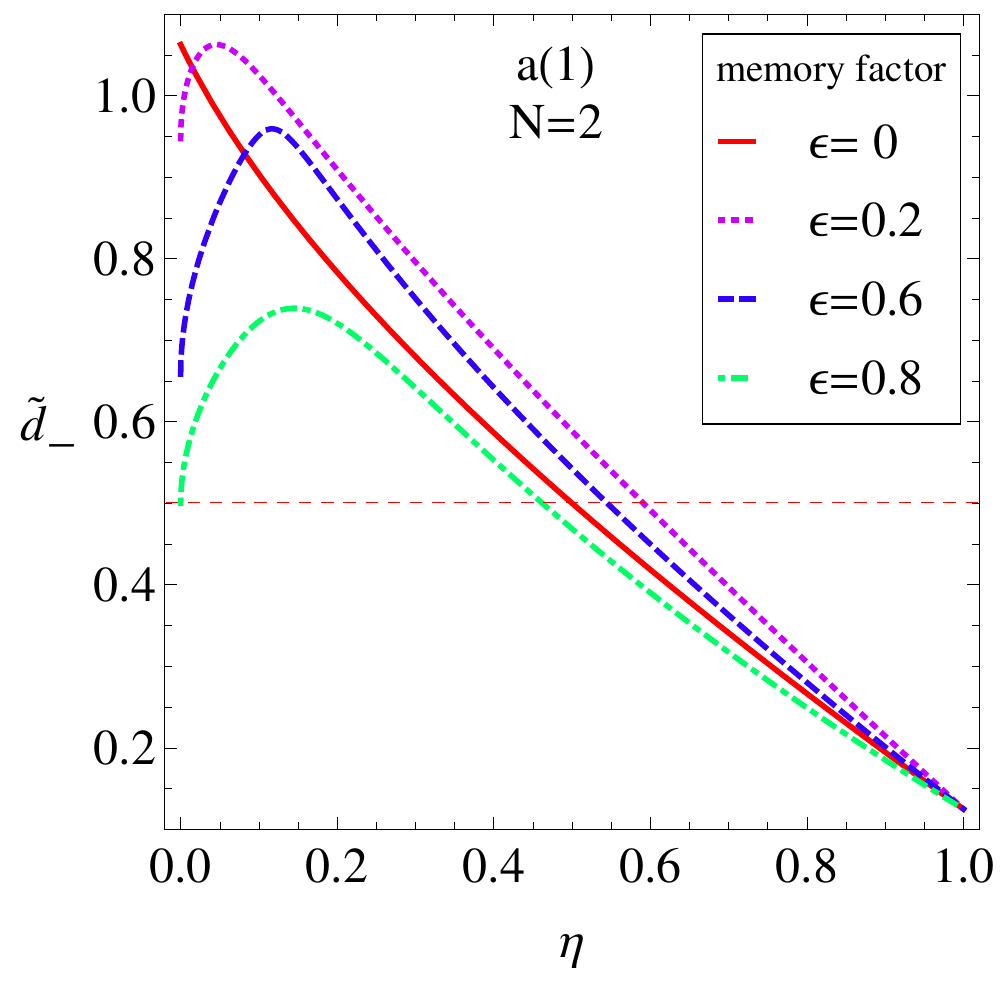}
\end{minipage}
\begin{minipage}[c]{0.45\textwidth}\centering
\includegraphics[width=3.3in]{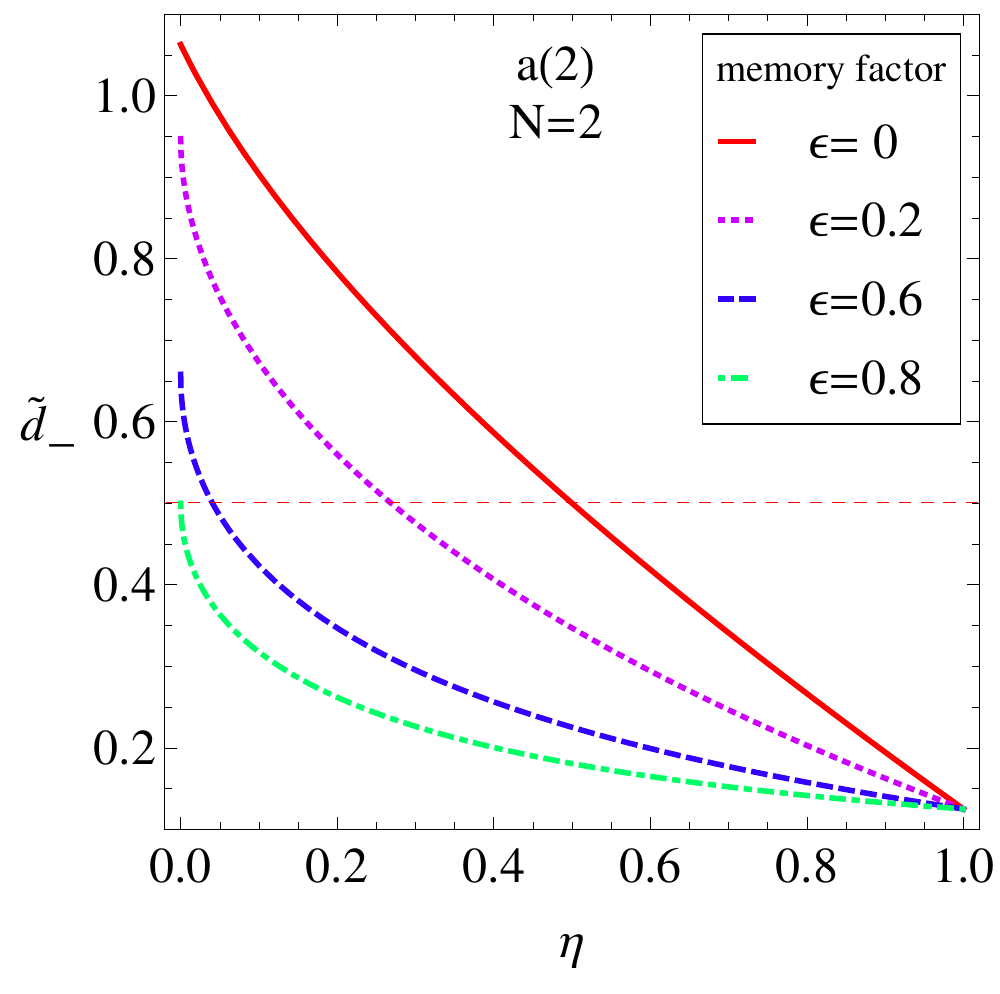}
\end{minipage}\\
\centering
\begin{minipage}[c]{0.45\textwidth}
\centering
\includegraphics[width=3.3in]{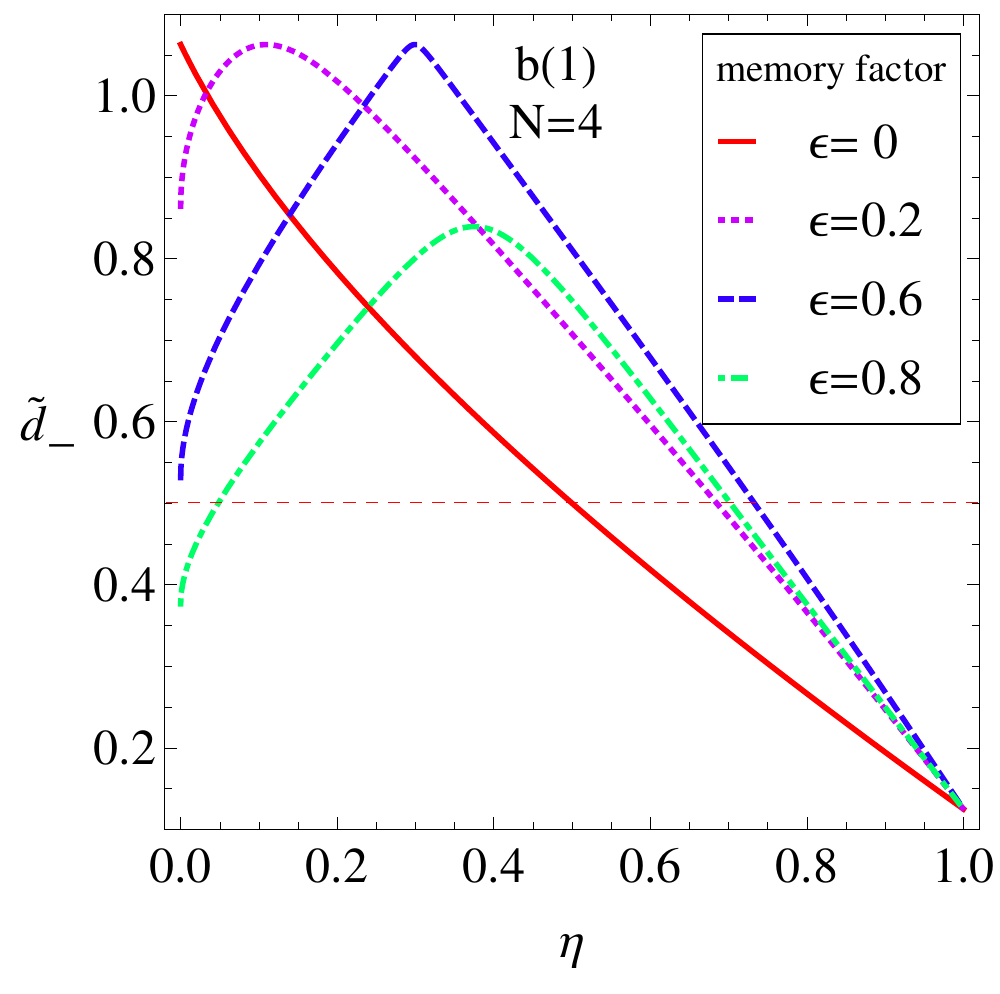}
\end{minipage}
\begin{minipage}[c]{0.45\textwidth}\centering
\includegraphics[width=3.3in]{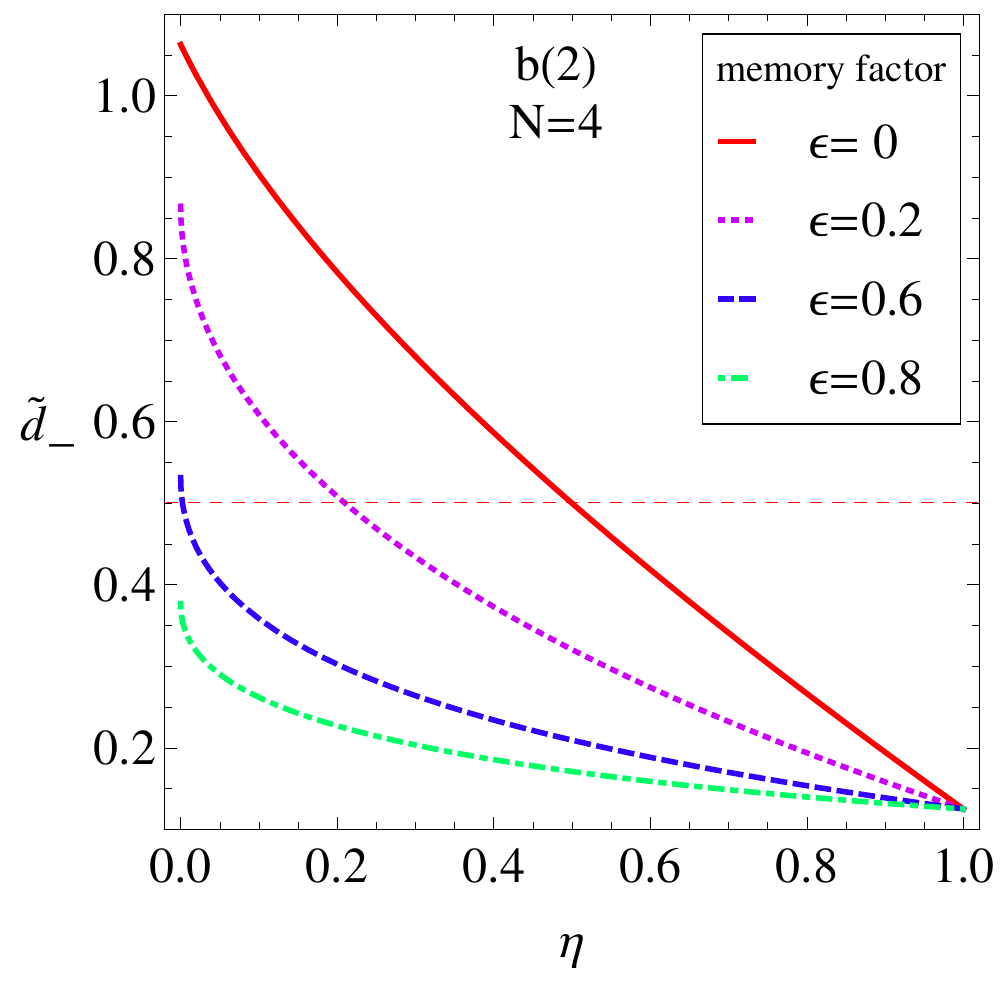}
\end{minipage}
\caption{(Color online)
The least symplectic eigenvalue $\tilde{d}_{-}$
of the partially transposed covariance matrix of the entangled states as a function of four parameters,
$\tilde{d}_{-}=\tilde{d}_{-}(\eta,\epsilon,T,\mu)$.
Phase flips have been implemented in figures a(2) and b(2), while in figures a(1) and b(1) they have not been taken.
We assume $T=1$ and $\mu =0.6$ in the numerical analysis. In order to facilitate the compare with the separability criterion,
$\tilde{d}_{-}=0.5$ is also shown as dashed lines.
Phase flips transform the memory factors become positive ingredients for
survival of entanglements when transmitting over the lossy bosonic
Gaussian memory channels.
}\label{entanglementsurvival}
\end{figure*}

The example state for testing the suppressing effects of our scheme is the coherent state
$|\alpha\rangle$, which is a Gaussian state characterized by its first moment and the covariance matrix
\begin{align}\label{coherent state}
    d_{\textmd{in}}&=(d_1,d_2)^\mathsf{T}=(\frac{\alpha+\alpha^{*}}{\sqrt{2}},\frac{\alpha-\alpha^{*}}{\sqrt{2}})^\mathsf{T},\\
    V_{\textmd{in}}&=\frac{1}{2}\mathbb{I}_{2 \times 2}.
\end{align}
The parameter $|\alpha|^2=\bar{n}$ expresses the average photon numbers of the mode flied. Without losing generality,
we assume the parameter $\alpha$ is real.

After some calculations, the numerical analysis of the fidelity \eqref{Gaussian fidelity}
can be given out as a function of four parameters: the transmissivity $\eta$, the memory factor $\epsilon$,
the average excitations $T$ and the average photon number $\alpha$, namely, $F=F(\eta, \epsilon, T, \alpha)$.
The numerical results for roles of the parameters are reported in Fig.~\ref{fidelity ps}.
As shown in a(1) and b(1), when phase flips are not being implemented, the memory factor produces a completely negative effect on
the signal transmission over the memory channels.
Interestingly, when phase flips are being implemented,
it becomes another important aspect for the improvement of fidelity which has been shown in a(2) and b(2).
The transform of the role of the memory factor from completely negative to positive is totally exciting.

Let's give a simple analysis for the extreme case as the transmissivity $\eta$ tends to zero
when no phase flips are being implemented. From Eq.~\eqref{memorycoefficientf} we have
\begin{equation}\label{tzero}
d_{\textmd{out}k} \simeq -\sqrt{\epsilon} d_{\textmd{in}k-1}-\sqrt{1-\epsilon} e_{k}+\sqrt{\epsilon} \delta_{k,1} m_1.
\end{equation}
For a perfect memory channel $\epsilon \rightarrow 1$, further, it becomes
$d_{\textmd{out}k}  \simeq  -d_{\textmd{in}k-1}+ \delta_{k,1} m_1$
which is a correlated dephasing channel. The phase error is determined exactly by whether an error occurred on the previous one.
For perfect memory channels, the facts is that they are asymptotically noiseless
where no information is lost to the environment~\cite{Bowen2004}.
This implies that for zero transmissivity channel, the fidelity maybe not zero. As the memory factor of the channel becomes large,
e.g. the perfect memory channel, the fidelity can be considerable since there are no information is lost.
This behaviors can be seen in Fig.~\ref{fidelity ps} a(1) and b(1) as the transmissivity reaches zero.
Such characteristic of memory also presents in the discussion of the survival of entanglements in Sec.~\ref{entanglements} ,
where the survival does not decrease monotonously with the decrease of the transmissivity,
see in Fig.~\ref{entanglementsurvival} a(1) and b(1).

Let's give a brief discussion to end this subsection.
Actually, gaining an insight one can find that a ``repetition coding'' has been structured in the encoding procedure of the scheme.
The first $N$-port splitter transforms the input coherent state into a product state
of $N$ coherent states with the same reduced amplitude: $|\beta\rangle^{\otimes N}=|\alpha/\sqrt{N}\rangle^{\otimes N}$
\cite{Andersen2013,Xiang2010}.
So, a train of signals of $N$ reduced coherent states is sent through the memory channel in sequences.
But this is not always the case, there exists such repetition coding. For example,
when only one half of a two-mode CV entangled state is sent through the channel,
there has certain complexity in analyzing the transformations of the input signal states for the first $N$-port splitter.
In the next subsection, we will investigate this in detail.
\begin{figure*}[htbp]
\centering
\begin{minipage}[c]{0.45\textwidth}
\centering
\includegraphics[width=3.3in]{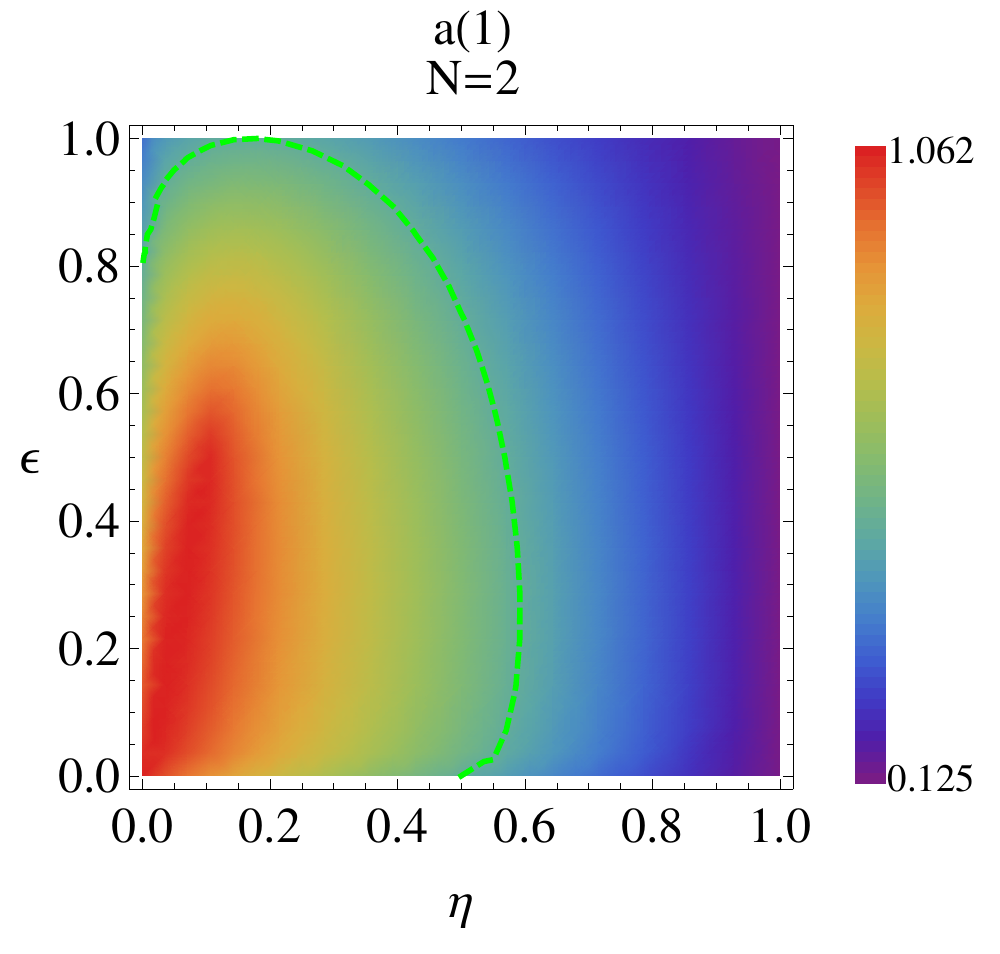}
\end{minipage}
\begin{minipage}[c]{0.45\textwidth}\centering
\includegraphics[width=3.3in]{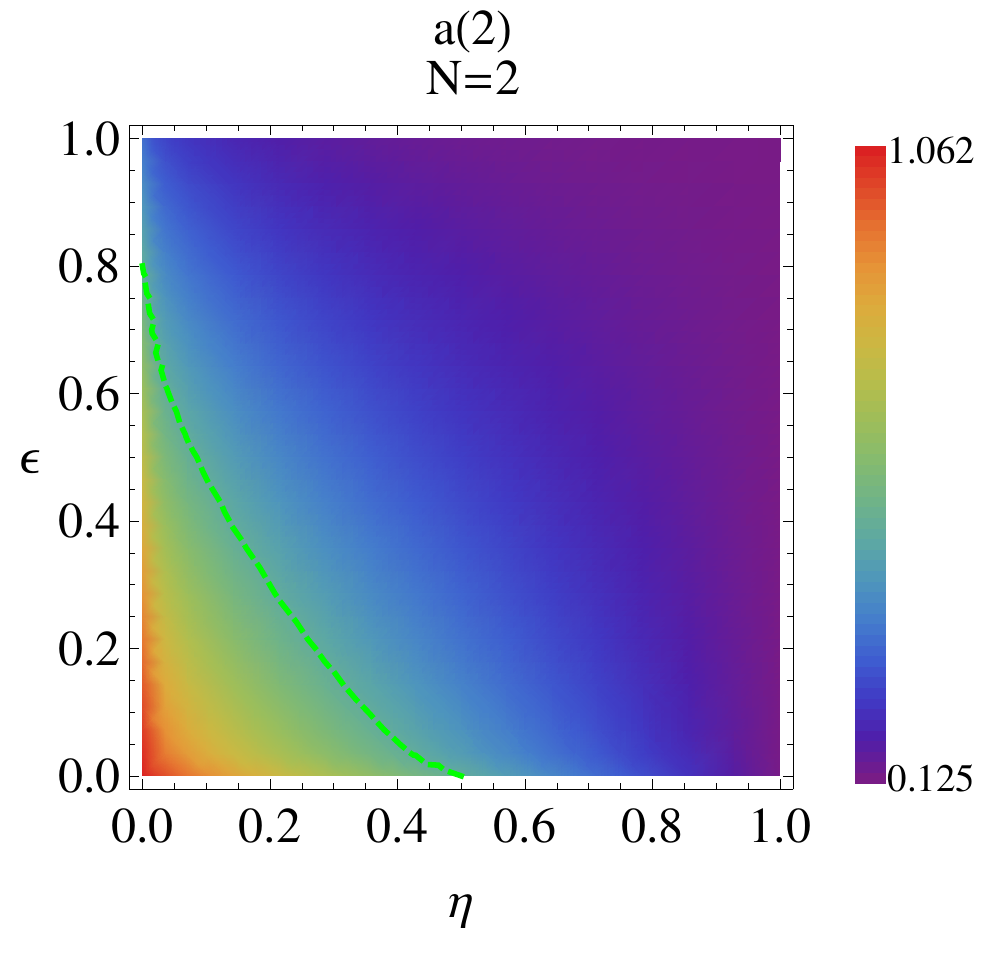}
\end{minipage}\\
\centering
\begin{minipage}[c]{0.45\textwidth}
\centering
\includegraphics[width=3.3in]{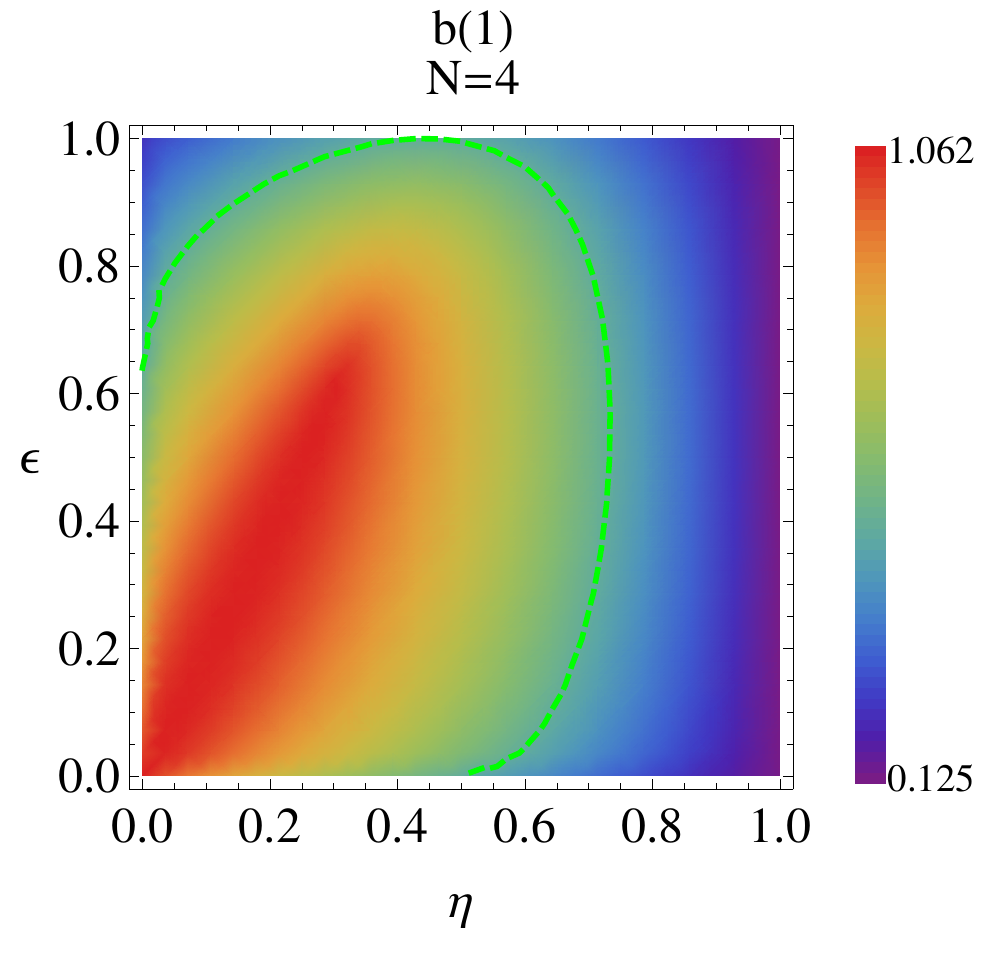}
\end{minipage}
\begin{minipage}[c]{0.45\textwidth}\centering
\includegraphics[width=3.3in]{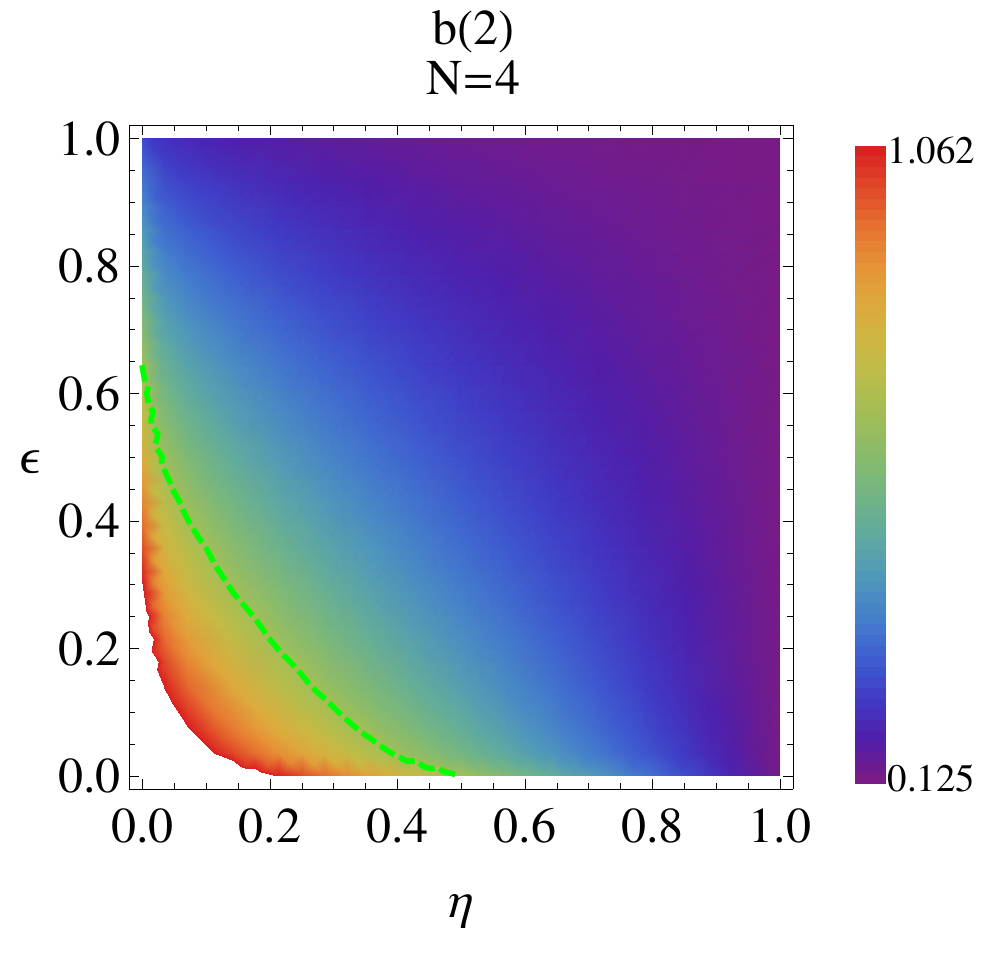}
\end{minipage}
\caption{(Color online)
The density plot of $\tilde{d}_{-}$ as a function of the four parameters: $\eta$, $\epsilon$, $T$, and $\mu$.
In figures a(2) and b(2) phase flips have been implemented , while in a(1) and b(1) they have not been taken.
We assume $T=1$ and $\mu =0.6$ in the numerical analysis. In order to facilitate the compare
with the separability criterion, the density $\tilde{d}_{-}=0.5$ is also shown as dashed lines in figures.
}\label{entanglementsDensity}
\end{figure*}

\subsection{The survival of entanglements}\label{entanglements}
We now investigate the survival of entanglements using our suppressing scheme.
The example state is TMSVs
\begin{equation}\label{TMSV}
|\textrm{TMSVs}\rangle=S(r)|00\rangle=\sqrt{1-\mu^2}\sum_{n=0}^{\infty}\mu^n|nn\rangle,
\end{equation}
where $\mu=\tanh r$ is the squeezing parameter. The TMSVs are two mode Gaussian states,
which are especially useful in the CV quantum information because they can be used as entanglement resources.
Considering the fundamental quantum teleportation protocol~\cite{Bennett1993,cvprotocol},
we let only one half of the TMSVs sending through the memory channel in the suppressing scheme.
The degree of the entanglement will be contaminated by the noise channel unavoidably.
Now, we can analyze the correlated noise suppressions for the survival of entanglements.

The covariance matrix of the TMSVs is a $4\times 4$ square matrix which can be written as
\begin{equation}\label{CM TMSV}
    V= \left(
                      \begin{aligned}
                        & A && C \\
                        &C^\mathsf{T} && B \\
                      \end{aligned}
                  \right),
\end{equation}
where $A=B=\cosh(2r) \mathbb{I}$ and $ C= \sinh(2r) \textmd{Diag}(1,-1)$.

One can let the mode $A$ transmit through the Gaussian memory channel,
so we have $A=V_{\textmd{in}}$. Furtherly, considering the
independent relations between the mode $B$ and the memory channel,
one can find the covariance matrix of output state is
\begin{equation}\label{CMoutTMSV}
    V'= \left(
                      \begin{aligned}
                        & A' && C' \\
                        &{C'}^\mathsf{T} && B' \\
                      \end{aligned}
                  \right),
\end{equation}
with $A'=V_{\textmd{out}}$, $B'=B$ and $C'=\zeta_{\textmd{in}} C$.
A detail calculation is reported in Appendix~\ref{calculation}.

A necessary and sufficient criterion exists for the separability of the two-mode bipartite
Gaussian states \cite{Simon2000,Duan2000,Serafini2005}:
a Gaussian state with the covariance matrix $V$ is separable iff $\tilde{d}_{-} > \frac{1}{2}$,
where $\tilde{d}_{-}$ is the least symplectic eigenvalue of the partially transposed
covariance matrix $\tilde{V}=\Delta V \Delta$ with $\Delta=\Det (1,-1,-1,-1)$. The symplectic
eigenvalues can be calculated from the spectrum of $i J \tilde{V}$~\cite{Ferraro2005}, where
\begin{equation}\label{J symplectic}
    J=\omega \bigoplus \omega, \hspace{3mm} \omega= \left(
                      \begin{aligned}
                        & 0 && 1 \\
                        &-1 && 0 \\
                      \end{aligned}
                  \right),
\end{equation}
or using the A. Serafini et al.'s~\cite{Serafini2004} formula directly.
Using the criterion, we can analyze the survival of entanglement of the TMSVs when transmitting over the noise memory channel.
The least symplectic eigenvalue $\tilde{d}_{-}$ can be calculated from
the covariance matrix $\tilde{V}=\Delta V' \Delta$ using the Eq.~\eqref{CMoutTMSV}, numerically.

Similarly with the fidelity of the coherent states transmit over the memory channel,
the $\tilde{d}_{-}$ relies on four parameters:
the transmissivity $\eta$, the memory factor $\epsilon$,
the average excitations $T$ for per environmental mode and the squeezing parameter $\mu$.
Numerical results shows that, before phase flips being implemented,
the survival of entanglement is complicatedly dependent on
both the parameters $\eta$ and $\epsilon$. While after phase flips being have implemented,
they all played important roles for maintaining the existence of entanglement transmitting over the memory channel,
as illustrated in Fig.~\ref{entanglementsurvival}.
The role of the memory factor in the scheme is always positive.
This also can be shown by aiding the density plot of $\tilde{d}_{-}$
for the parameters $\eta$ and $\epsilon$ in Fig.~\ref{entanglementsDensity}.
Increasing the numbers of beam splitters also plays
positive roles in maintaining the existence of entanglement,
which also can be shown in figures by comparing the cases $N=2$ and $N=4$.

Once again, the phase flips transform the memory factor as a positive ingredient for
the survival of entanglement when transmitting over such lossy bosonic memory channels.
By aiding the positive roles of the memory factor in the scheme,
the correlated noise are being greatly suppressed and thus the input quantum states are exhibited excellent protections.
\section{Summary}\label{summary}
In summary, we have proposed a scheme for suppressing the correlated noise in signals transmitted
over the Gaussian memory channels. The scheme is based on linear optical elements,
two $N$-port splitters and $N$ number of phase flips.
This is a compromise scheme which suppress rather than remove the noise,
but no predetermined condition of the correlated noise is required.

The suppressing efficiency of the correlated noise has been examined
for quantum information both from
quantum states directly transmitted over
the noise channel and also from the entanglement teleportation.
The proposed scheme demonstrates promising advantages, where the correlated noise of the memory channels are greatly suppressed,
and the input signal states are excellent protected when transmitting over the noise channels.

Numerical results show that
the phase flips are very important operations for the suppressions of the correlated noise.
They transform the roles of the memory channel from completely negative to positive in quantum communication.
Increasing the numbers of beam splitters is also helpful in reducing the noise of the quantum communication.

A general analysis beyond a specific lossy Gaussian memory channel mode is interesting.
In the suppressing scheme, the improvement of information capacities of
noise channels with correlated noise is also interesting.

\appendix
\section{Detail calculation}\label{calculation}
Defining $R:=(q_{\textmd{a}}, p_{\textmd{a}}, q_{\textmd{b}}, p_{\textmd{b}})^\mathsf{T}$,
$R_{\textmd{a}}:=(q_{\textmd{a}}, p_{\textmd{a}})^\mathsf{T}$ and
$R_{\textmd{b}}:=(q_{\textmd{b}}, p_{\textmd{b}})^\mathsf{T}$,
since only the mode $A$ has transmitted through the channel while the other did not,
we have $R_{\textmd{a}}=R_{\textmd{in}}$, $R'_{\textmd{a}}=R_{\textmd{out}}$ and $R'_{\textmd{b}}=R_{\textmd{b}}$.
According to the definition of the covariance matrix~\cite{Ferraro2005}
\begin{equation}\label{CMCout00}
    V'_{jk}:=\frac{1}{2}\langle \{R'_{j},R'_{k}\}\rangle
             - \langle R'_{j}\rangle  \langle R'_{k}\rangle,
\end{equation}
the relations $A'=V_{\textmd{out}}$ and $B'=B$ in \eqref{CMoutTMSV}  are easily understandable.

And also, according to the definition of the covariance matrix,
the form of the matrix $C'$ can be obviously written as
\begin{equation}\label{CMCout01}
    C'_{jk}:=\frac{1}{2}\langle \{R_{\textmd{out}j},R_{\textmd{b}k}\}\rangle
             - \langle R_{\textmd{out}j}\rangle  \langle R_{\textmd{b}k}\rangle.
\end{equation}
Noting that the mode $B$ did not transmit through the channel, so it is independent with the auxiliary vacuum states,
the initial memory mode and the environment of the memory channel.
Thus we have
\begin{align}\label{CMCout01}
    C'_{jk}&=\zeta_{\textmd{in}}\left[\frac{1}{2}\langle \{R_{\textmd{in}j},R_{\textmd{b}k}\}\rangle
             -\langle R_{\textmd{in}j}\rangle  \langle R_{\textmd{b}k}\rangle\right]\nonumber\\
           &=\zeta_{\textmd{in}}C_{jk},
\end{align}
where we have used the Eq.~\eqref{channelR} to connect the output with the input operators.

\begin{center}
\textbf{ACKNOWLEDGMENTS}
\end{center}

We would like to thank Cosmo Lupo and Ulrik Lund Andersen for helpful
comments. This work is supported by 973 Program under Grant No. 2010CB922904,
the National Natural Science Foundation of China under
Grant Nos. 11175248 and 61072034, and partly supported by the Basic Research
Foundation of Engineering University of CAPF (WJY-201104). S.-M. Ke is supported by
the Shaanxi Province Natural Science Foundation Research Project under Grant No. 2013JQ1011, and
the Special Fund for Basic Scientific Research of Central
Colleges, Chang'an University and the Special Foundation for Basic
Research Program of Chang'an University  under Grant No. CHD2012JC019.

\newcommand{\PRL}{Phys. Rev. Lett. }
\newcommand{\PRA}{Phys. Rev. A }
\newcommand{\JPA}{J. Phys. A }
\newcommand{\JPB}{J. Phys. B }
\newcommand{\PLA}{Phys. Lett. A }


\begin{thebibliography}{100}
\bibitem{Bennett1993} C. H. Bennett, G. Brassard, C. Crepeau, R. Jozsa, A. Peres, and W. K. Wootters, \PRL {\bf 70}, 1895 (1993).

\bibitem{Gisin2002} N. Gisin, G. G. Ribordy, W. Tittel, and H. Zbinden, Rev. Mod. Phys. {\bf 74}, 145 (2002).

\bibitem{cvprotocol} L. Vaidman, \PRA {\bf 49}, 1473 (1994);
                     S. L. Braunstein, and H. J. Kimble, \PRL {\bf 80}, 869 (1998);
                     T. C. Ralph, and P. K. Lam, \PRL {\bf 81}, 5668 (1998).

\bibitem{cvrmp}     C. Weedbrook, S. Pirandola, R. Garcia-Patron, N. J. Cerf, J. H. Shapiro, and S. Lloyd,
                    Rev. Mod. Phys. {\bf 84}, 621 (2012).

\bibitem{qec} S. L. Braunstein, Nature {\bf 394}, 47 (1998);
              T. Aoki, et al., Nature Phys. {\bf 5}, 541 (2009);
              D. Gottesman, arXiv:0904.2557;
              T. B. Pittman, B. C. Jacobs, and J. D. Franson, \PRA {\bf 71}, 052332 (2005);
              M. Lassen, et al., Nature Phot. {\bf 4}, 700 (2010).

\bibitem{Bowen2005} G. Bowen, I. Devetak and S. Mancini, \PRA {\bf 71} 034310 (2005);

\bibitem{Corney2006} J. F. Corney, P. D. Drummond, J. Heersink, V. Josse, G. Leuchs, and U. L. Andersen, \PRL{\bf 97}, 023606 (2006).

\bibitem{Kretschmann2005} D. Kretschmann and R. F. Werner, \PRA{\bf 72}, 062323 (2005).

\bibitem{Caruso2012} F. Caruso, V. Giovannetti, C. Lupoz, S. Mancini, arXiv:1207.5435.

\bibitem{memorycodes} J. P. Clemens, S. Siddiqui, and J. Gea-Banacloche, \PRA {\bf 69}, 062313 (2004);
                      R. Klesse and S. Frank, \PRL {\bf 95}, 230503 (2005);
                      A. D¡¯Arrigo, E. De Leo, G. Benenti, and
                      G. Falci, Int. J. Quantum Info. {\bf 6}, 651 (2008);
                      C. Cafaro and S. Mancini, Phys. Lett. A {\bf 374}, 2688 (2010); \PLA{\bf 82}, 012306 (2010).

\bibitem{Datta2007} N. Datta and T. C. Dorlas, J. Phys. A {\bf 40}, 8147 (2007).

\bibitem{Arrigo2007} A. D' Arrigo, G. Benenti, and G. Falci, New J. Phys. {\bf 9}, 310 (2007).

\bibitem{Giovannetti2005} V. Giovannetti, J. Phys. A {\bf 38}, 10989 (2005).

\bibitem{Lupo2012} C. Lupo, L. Memarzadeh, and S. Mancini, \PRA {\bf 85}, 012320 (2012).

\bibitem{Lassen2013} M. Lassen, A. Berni, L. S. Madsen, R. Filip, and U. L. Andersen, \PRL {\bf 111}, 180502 (2013).

\bibitem{magnitude} K. Banaszek, A. Dragan, W. Wasilewski, and C. Radzewicz, \PRL {\bf 92}, 257901 (2004);
                   R. Demkowicz-Dobrza\'{n}ski, P. Kolenderski, K. Banaszek, \PRA {\bf 76}, 022302 (2007).

\bibitem{2Nsplitter} K. Mattle \textit{et al}., Appl. Phys. B {\bf 60}, S111 (1995);
                     P. T\"{o}rm\"{a} and I. Jex, J. Mod. Opt. {\bf 43}, 2403-2408 (1996);
                     M. \.{Z}ukowski, A. Zeilinger, M. A. Horne, \PRA{\bf 55}, 2564 (1997);
                     G. J. Pryde, A. G. White, \PRA {\bf 68}, 052315 (2003);
\bibitem{Lupo2010} C. Lupo, V. Giovannetti, and S. Mancini, \PRL {\bf 104}, 030501 (2010);
                                                             \PRA {\bf 82}, 032312 (2010).

\bibitem{Ferraro2005} A. Ferraro, S. Olivares, M. G. A. Paris, arXiv:quant-ph/0503237.

\bibitem{LupoMancini2010} C. Lupo, S. Mancini, \PRA {\bf 81}, 052314 (2010).

\bibitem{Holevo2001} A. S. Holevo and R. F.Werner, \PRA {\bf 63}, 032312 (2001).

\bibitem{Scutaru1998} H. Scutaru, \JPA {\bf 31}, 3659 (1998).

\bibitem{Meisheng2007} M. Zhao, T. Qin, Y. Zhang, Mod. Phys. Lett. B {\bf{21}}, 1531 (2007).

\bibitem{Isara2008} A. Isara, Eur. Phys. J. Special Topics {\bf{160}}, 225-234 (2008).

\bibitem{Marian2012} P. Marian and T. A. Marian, \PRA {\bf 86}, 022340 (2012).

\bibitem{Bowen2004} G. Bowen and S. Mancini, \PRA{\bf 69}, 012306 (2004).

\bibitem{Andersen2013} U. L. Andersen and T. C. Ralph, \PRL {\bf 111}, 050504 (2013).

\bibitem{Xiang2010} G. Y. Xiang, T. C. Ralph, A. P. Lund, N. Walk and G. J. Pryde, Nat. Photonics {\bf 4}, 316-319 (2010).

\bibitem{Simon2000} R. Simon, \PRL {\bf 84}, 2726 (2000).

\bibitem{Duan2000} L.-M. Duan, G. Giedke, J. I. Cirac, and P. Zoller, \PRL {\bf 84}, 2722 (2000).

\bibitem{Serafini2005} A. Serafini, M. G. A. Paris, F. Illuminati and S. De Siena, J. Opt. B {\bf 7}, R19-R36 (2005).

\bibitem{Serafini2004} A. Serafini, F. Illuminati and S. De Siena, J. Phys. B: At. Mol. Opt. Phys. {\bf 37}, L21-L28 (2004).

\end{thebibliography}
%


\end{document}